\documentclass[twocolumn,fleqn,showpacs,showkeys]{revtex4-1}
\usepackage{graphicx}
\usepackage{amssymb}
\usepackage{amsmath}
\usepackage{bm}
\usepackage{hyperref}
\begin{document}

\title{Time Reversibility of Quantum Diffusion in Small-world Networks}
\author{Sung-Guk Han}
\author{Beom Jun Kim}
\email[Corresponding author: ]{beomjun@skku.edu}
\affiliation{Department of Physics and BK21 Physics Research Division, Sungkyunkwan University, Suwon 440-746, Korea}

\date{\today}

\begin{abstract}
We study the time-reversal dynamics of a tight-binding electron in the
Watts-Strogatz (WS) small-world networks.  The localized initial wave packet at
time $t=0$ diffuses as time proceeds until the time-reversal operation,
together with the momentum perturbation of the strength $\eta$, is made at the
reversal time $T$.  The time irreversibility is measured by $I \equiv |\Pi(t =
2T) - \Pi(t = 0)|$, where $\Pi$ is the participation ratio gauging the extendedness
of the wavefunction and for convenience, $t$ is measured forward even after the time reversal
. When $\eta = 0$, the time evolution after $T$ makes the
wavefunction at $t=2T$ identical to the one at $t=0$, and we find $I=0$,
implying a null irreversibility or a complete reversibility.  On the other
hand, as $\eta$ is increased from zero, the reversibility becomes weaker, and we
observe enhancement of the irreversibility.  We find that
$I$ linearly increases with increasing $\eta$ in the weakly-perturbed region, and that
the irreversibility is much stronger in the WS network than in the local
regular network.
\end{abstract}

\pacs{89.75.Hc, 73.20.Jc, 05.70.Ln}
\keywords{Small-world network, Quantum diffusion, Time-reversal dynamics}

\maketitle 
Time-irreversible phenomena in nature, such as spread of an ink blob in water
and aging of living organisms, are ubiquitous in our macroscopic 
world~\cite{Lebowitz}.
We always see the ink blob spread in a teacup and people get older, not
the other way around. These time-irreversible behaviors look 
puzzling because all these dynamics are based on microscopic equations of
motion, quantum or classical, that contain time-reversal symmetry.
In the late 19th century, this seemingly paradoxical observation of 
time-reversal microscopic dynamics and time-irreversible macroscopic
thermodynamics perplexed many scientists until a breakthrough was
made by Boltzmann. It is now well-known that the directionality of {\it the
arrow of time} in the second law of thermodynamics can be understood by the
huge difference in the numbers of allowed microscopic states between the initial
and the final macroscopic states.

Recently, the quantum and the classical time-reversal dynamics of 
a diffusion system and a spin system have been studied to answer 
the long-standing question of where
does the irreversibility come from~\cite{Petitjean,Yamada,Hiller,Waldherr}. In
Ref.~\onlinecite{Yamada}, quantum diffusion systems  were shown to
display the universal behavior of time irreversibility in terms of perturbation
strength. Because the complex networks have drawn much interest, there
have been studies of quantum systems in Watts-Strogatz (WS) 
small-world networks~\cite{Watts}. 
Localization phenomena of quantum systems in small-world networks have
been studied~\cite{Giraud,Chenping}, 
and quantum diffusion problems have been investigated in
various ways in WS networks~\cite{Monasson,Kim,Mulken1,Mulken2}.

In the present work, we aim to study the time reversibility of the quantum
diffusion problem in WS networks.
The diffusion of a wave packet has been shown to occur much faster in WS networks 
than in a regular network~\cite{Kim,Mulken1}: A suitably
defined diffusion time in a quantum system shows $\tau \sim \log N$ in the former while $\tau \sim
N$ for the latter structure. An abrupt change in
the scaling behavior of $\tau$ has been shown to occur as soon as the rewiring probability
becomes nonzero. In other words, a transition from the slow world ($\tau \sim
N$) to the fast world ($\tau \sim \log N$) occurs at null rewiring probability
simultaneously with a structural transition from the large world ($l \sim N$)
to the small world ($l \sim \log N$), with $l$ being the average path length
connecting two arbitrarily chosen vertices. Consequently, a change in the
scaling behavior of $\tau$ reflects a dynamical aspect of the
small-world transition in WS networks. 
In this work, we examine the effect of shortcuts on the time reversal dynamics of 
the tight-binding electron system at different perturbation strengths. 

The time evolution of the  tight-binding electron in a network structure is
governed by the following time-dependent Schr\"odinger equation:
\begin{equation}
\label{eq:Sch}
i \hbar \frac{\partial\vert\Psi\rangle}{\partial t} = H\vert \Psi \rangle, 
\end{equation} 
where $\hbar$ is the Planck constant, and 
the quantum mechanical ket $\vert\Psi \rangle$ and the 
Hamiltonian $H$ in position representation are written as  
$\Psi_{n} \equiv \langle n\vert \Psi\rangle$  and 
\begin{equation}
\label{eq:H}
H_{n n'}=H_{n' n}=\left\{ \begin{array}{cl}
\Delta & \textrm{for } n'\in \Lambda_{n} \\
0 & \textrm{otherwise,}
\end{array}\right.
\end{equation}
respectively.
Here, $n$ is the vertex index, $\Lambda_n$ is the set of directly
connected vertices of $n$, and the on-site energy has been assumed 
to be uniform and set equal to zero. As a further simplification,
we also assume that the hopping energy $\Delta$ does not depend on 
$n$ or $n'$. Henceforth, we use the dimensionless units:
the time is measured in units of $\hbar/\Delta$, the position
in units of the lattice spacing $a$ of the one-dimensional (1D)
regular lattice without shortcut, and the momentum in units
of $\hbar/a$. 

Once the WS network for a given rewiring probability $p$ is constructed
following the standard procedure (see Ref.~\onlinecite{Watts}),  we 
perform the numerical integration of the time-dependent 
Schr\"odinger equation  starting from 
the initial localized wave packet at time $t=0$ ($\Psi_n = \delta_{n,0}$) 
by using the fourth-order Runge-Kutta algorithm
with the discrete time step $dt = 0.01$. Except for $p=0$, all results
are obtained from the average of 1000 different network realizations.
We use the {\it time-reversal} test similarly to Ref.~\onlinecite{Yamada}:
At the reversal time $T$, a momentum perturbation of the strength $\eta$ 
is made by applying the operator $\exp(i\eta\hat{x})$ to 
$| \Psi (T) \rangle$ with the position operator $\hat{x}$.
For convenience, we measure $t$ in the forward direction even after 
the reversal time $T$ so that at $t = 2T$, the system goes back to the 
initial state in the absence of the perturbation (i.e., when $\eta = 0$).
During the numerical time integration, we compute the participation ratio
\begin{equation}
\label{eq:Pi}
\Pi(t)=\frac{\sum_{n=1}^{N}{|\Psi_{n}(t)|^{2}}}{\sum_{n=1}^{N}|\Psi_{n}(t)|^4}, 
\end{equation}
which has been frequently used in the study of localization phenomena.
When the wavefunction is completely localized, $\Pi$ takes the value of unity.
On the other hand, as the wavefunction spreads in space and the quantum 
system becomes extended, $\Pi$ is known to be $O(N)$.
The time irreversibility $I$ is then defined as
\begin{equation}
\label{eq:I}
I \equiv \vert \Pi(2T)- \Pi(0) \vert.
\end{equation}
It is to be noted that $\Pi(t < T)$ is independent of $\eta$ whereas 
$\Pi(2T)$ and $I$ are functions of $\eta$. When $\eta = 0$, 
$I$ takes the value zero, meaning the null irreversibility (or the
complete reversibility), because $\Pi(2T) = \Pi(0)$. As $\eta$ is increased
from zero, the perturbed momentum at $T$  makes it difficult for the system to
go back to the initial quantum state; thus, $I$ is expected to increase.

\begin{figure}
\includegraphics[width=0.48\textwidth]{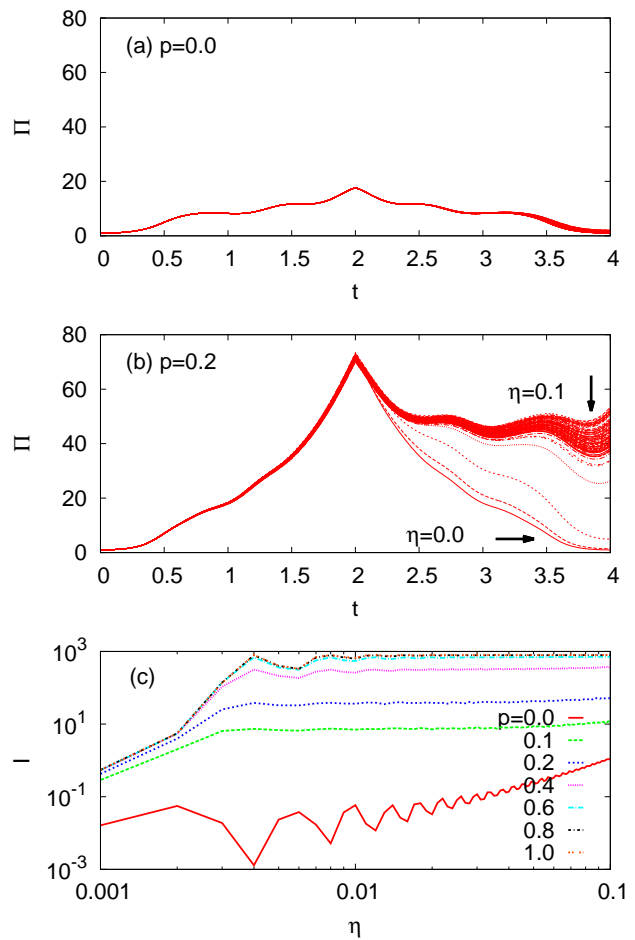}
\caption{(Color online)
The participation ratio $\Pi(t)$ as a function of time $t$ 
in (a) the local regular network and (b) the WS network at
rewiring probability $p=0.2$. 
The system size $N=1600$ and the reversal time $T=2$ are used.
Although not clearly discernible, (a) and (b) contain 101 different
curves for the momentum perturbation strength $\eta = 0.0, 0.001, 0.002, 
\cdots, 0.1$ (from bottom to top). The perturbed and time-reversed quantum
state in the WS network in (b) shows a much stronger deviation from the initial
localized quantum state.
(c) The irreversibility $I$=$|\Pi(2T)-\Pi(0)|$ versus $\eta$ at 
various rewiring probabilities $p$. When $p=0.0$ for the local regular network,
$I$ increases with increasing $\eta$. On the contrary, when $p\neq 0.0$,
$I$ first increases, but soon saturates. The oscillatory behavior observed
for $p=0.0$ originates from the discreteness of the underlying periodic
lattice structure.
}
\label{fig1}
\end{figure}

Figure~\ref{fig1} displays $\Pi(t)$ in Eq.~(\ref{eq:Pi}) (a) for the local
1D regular network corresponding to the WS network at the rewiring
probability $p=0$ and (b) for the WS network at $p=0.2$. Although not clearly
discernible, each of Fig.~\ref{fig1}(a) and (b) has 
101 different curves corresponding
to $\eta = 0.0, 0.001, 0.002, \cdots, 0.1$ (from bottom to top). 
We also show the time
irreversibility $I$ at various values of $p$ in Fig.~\ref{fig1}(c). The system
size $N=1600$ and the reversal time $T=2$ are used. We first observe that
$\Pi(T)$ is about four times larger for $p=0.2$ than for $p=0.0$, which is
explained by the fast diffusion in the WS network structure, as discussed in
Ref.~\onlinecite{Kim}.
At both (a) $p=0$ and (b) $p=0.2$, time reversal dynamics is shown to give rise to a larger 
reflection asymmetry of $\Pi(t)$ around $t=T$ as $\eta$ is increased.
This is not surprising because a larger momentum perturbation will
surely make the time reversed quantum state at $t=2T$ more different from
the initial state at $t=0$. A more interesting observation one can make
from the comparison of Fig.~\ref{fig1} (a) and (b) is that the deviation
of $\Pi(2T)$ from $\Pi(0)$ is much bigger for the WS network ($p=0.2$) 
than for the regular network ($p=0.0$). This clearly indicates that
the structural irregularity in the WS network plays an important role
in making the perturbed dynamics deviate severely from the unperturbed one.
The enhancement of the irreversibility in the WS network is more clearly
seen in Fig.~\ref{fig1} (c) for $I$ versus $\eta$ at various rewiring
probabilities $p$. Figure~\ref{fig1}(c) shows that the irreversibility 
is a monotonically increasing function of the rewiring probability $p$.
In other words, the enhanced structural randomness caused by more shortcuts
drives the quantum diffusion dynamics in the WS network to become more sensitive
to the momentum perturbation. Interestingly, the
WS network at a sufficiently larger rewiring probability behaves very 
differently from the local regular lattice: For $p > 0$, 
$I$ increases linearly with increasing $\eta$ for the weakly-perturbed 
region and then soon saturates to a finite value. In contrast, for $p=0$, $I$ increases 
linearly with increasing $\eta$ without showing saturation. 
However, these findings need to be carefully
examined in view of the finite scales in the system, i.e., finiteness of 
the length scale ($N$) and the time scale ($T$).

\begin{figure}
\includegraphics[width=0.48\textwidth]{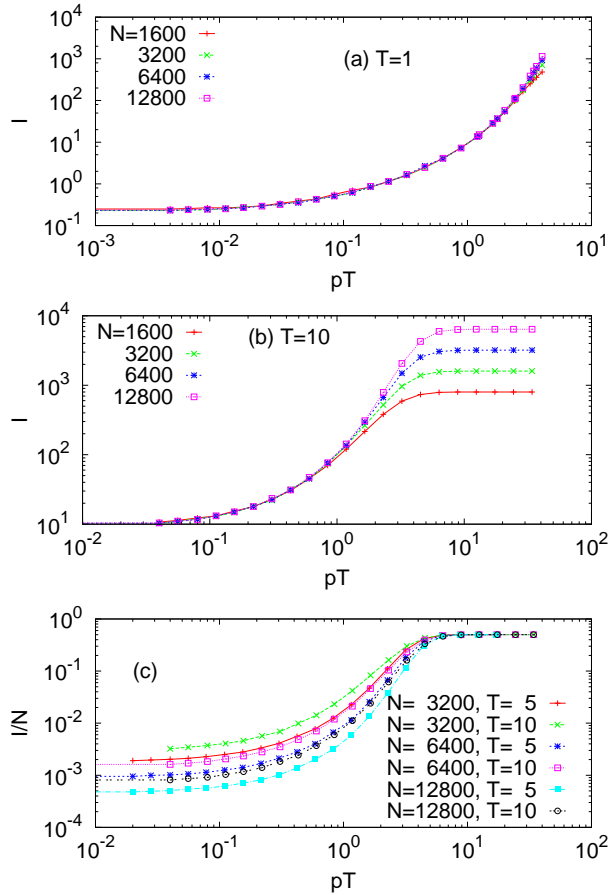}
\caption{(Color online) The time irreversibility $I$ in Eq.~(\ref{eq:I}) is measured
for various values of the rewiring probability $p$ at a fixed value
of the reversal time $T$ to plot $I$ versus $pT$ for (a) $T=1$
and (b) $T=10$. For $pT \lesssim 1$, the system is in the 1D  region
and $I$ does not depend on the system size $N$. As $pT$ is increased
to become $pT \gtrsim 1$, the system enters  the small-world region
and different sizes begin to show deviations from each other.
As $pT$ is increased further, $I$ saturates to some finite value:
$I(pT \rightarrow \infty)/N \sim$ const., independent of $T$, as shown
in (c).
The strength of momentum perturbation is $\eta=0.1$, and all results are
averages over 1000 different network realizations.
} 
\label{fig2}
\end{figure}

It should be noted that in WS networks there is an additional length
scale other than the microscopic length scale of the lattice spacing $a$
and the macroscopic length scale of the system size $N$. When a nonzero
rewiring probability $p$ is given, the number of shortcuts is proportional to
$pN$, which determines the third length scale of the average distance $\xi$
between the two endpoints of shortcuts, i.e., $\xi \sim N/pN = 1/p$.
If the tail part of the wavefunction of the tight-binding electron does not
have enough time to arrive at the closest shortcut endpoint, the system should
behave just like a 1D  regular lattice. 
Also, quantum diffusion in a regular 1D system is known to have a
diffusion time proportional to the system size to become fully extended ($\tau \sim N$), 
which indicates that the wavepacket in the 1D system spreads on a distance
scale $\ell \sim T$ before reversal time. Accordingly, 
if $T \sim \ell \lesssim 1/p$, the system behaves just like a
1D regular lattice while it changes its behavior as $T \sim 1/p$ is crossed.
Let us consider several cases: (i) For $pT \lesssim 1$, the system behaves
like a 1D system because $T$ is short enough that the wavepacket does
not have time to arrive at a shortcut endpoint. 
Furthermore, in this case of short-time diffusion, all observed results 
should be independent of $N$ as long as $T \ll N$; i.e., $T$ is too short
for the particle to feel the finiteness of the system. (ii) For $pT \gtrsim 1$,
the tight-binding particle now begins to meet shortcut endpoints and the
diffusion behavior changes. As soon as the
WS network begins to have a finite fraction of shortcuts, the quantum 
diffusion behavior is known to rapidly change so that the diffusion time $\tau$
increases only logarithmically with increasing $N$. This implies that the wavefunction
spreads in a distance that increases exponentially with time. We, thus,
conclude that this very fast diffusion occurs when $pT \gtrsim 1$ and
that the particle can arrive at the other side of the system, giving rise
to an $N$-dependence in the irreversibility. Consequently, the 
size-independence of $I$ in the region of $pT \lesssim 1$ is changed as
we enter the intermediate region of $pT \gtrsim 1$. (iii) For the limiting
case of $pT \gg 1$, the wavefunction of the particle has already completely
spread; thus, the irreversibility should not depend on $pT$. In this
long-time limit, the wavefunction becomes fully extended, and 
the participation ratio is $\Pi \sim N$. Consequently, for a big enough system,
$I(pT)/N$ saturates as $pT$ becomes larger, approaching a value that
does not depend on $N$.

The results  for the irreversibility $I$ versus $pT$ curve at $\eta = 0.1$ 
shown in Fig.~\ref{fig2} clearly fit the above hand-waving scaling arguments well.
In Figs.~\ref{fig2}(a) and (b), $I$ is shown to be independent of $N$ 
for $pT \lesssim 1$ whereas we see a clear size dependence for $pT \gtrsim 1$.  
As $pT$ is increased even further, $I$ saturates to some values that depend on $N$, 
as clearly seen in Fig.~\ref{fig2}(b). Figure~\ref{fig2}(c) shows that the saturated 
value of $I/N$ in the limit of $pT \gg 1$ is independent of $N$.  

In summary, we studied the time reversal dynamics of the tight-binding 
electron system in the WS small-world networks. Initially, the localized quantum 
mechanical state of the electron evolves in time until the momentum perturbation 
of the strength $\eta$ is made at $T$, at which time reversal operation is also made. 
The 1D regular lattice and the WS small-world network exhibit different behaviors: 
the irreversibility, measured by the difference between the participation ratio of 
the final and the initial states, is found to be bigger in the latter network 
due to the structural randomness caused by the random shortcuts.  
It is also found and argued that the irreversibility does not depend on the system 
size for $pT \lesssim 1$ and that it saturates to a value as $pT$ becomes larger.  

\acknowledgments
This work was supported by the Korea Research Foundation Grant
KRF-2009-013-C00021.  

\end{document}